\documentclass[aps,prl,twocolumn,preprintnumbers,superscriptaddress,amsmath,amssymb,footinbib]{revtex4-2}
\usepackage{graphicx,epsfig}
\usepackage{bm}
\usepackage{dcolumn}
\usepackage[english]{babel}
\usepackage[utf8]{inputenc}
\usepackage{comment}
\usepackage{amsmath}
\usepackage{scalerel}
\usepackage{lipsum}
\usepackage{amsfonts}
\usepackage{amssymb}
\usepackage{color}
\usepackage[usenames,dvipsnames]{xcolor}
\usepackage[colorlinks,bookmarks=false,citecolor=NavyBlue,linkcolor=Red,urlcolor=blue,
]{hyperref}
\usepackage[export]{adjustbox}
\usepackage{amsthm}
\usepackage{simplewick}
\usepackage{bbold}
\usepackage{ wasysym }
\usepackage[title]{appendix}
\usepackage{mathrsfs}  
\usepackage{braket}
\usepackage{graphics}
\usepackage{todonotes}

\usepackage{subfigure}

\usepackage{multirow}
\usepackage{algorithm}
\usepackage{algpseudocode}

\newcommand{\be}{\begin{equation}}
\newcommand{\ee}{\end{equation}}
\newcommand{\bea}{\begin{eqnarray}}
\newcommand{\eea}{\end{eqnarray}}

\newcommand{\Id}{\mathbb{1}}
\newcommand{\nsamp}{\mathcal{N}}  
\newcommand{\Trace}{\text{Tr}}

\newcommand{\Var}{\text{Var}}

\newcommand{\Pn}{q_n}   
\newcommand{\PnEx}{\tilde{q}_n}
\newcommand{\PnOne}{q_{1}}  
\newcommand{\PnExOne}{\tilde{q}_{1}}

\date{ \today} 
\begin{document}

\title{Quantum Magic via Perfect  Pauli Sampling of Matrix Product States}
\author{Guglielmo Lami}
\affiliation{International School for Advanced Studies (SISSA), 34136 Trieste, Italy}
\author{Mario Collura}
\affiliation{International School for Advanced Studies (SISSA), 34136 Trieste, Italy}
\affiliation{INFN Sezione di Trieste, 34136 Trieste, Italy}

\begin{abstract}
We introduce a novel breakthrough approach to evaluate the nonstabilizerness of an $N$-qubits Matrix Product State (MPS) with bond dimension $\chi$. In particular, we consider the recently introduced Stabilizer R\'enyi Entropies (SREs). We show that the exponentially hard evaluation of the SREs can be achieved by means of a simple perfect sampling of the many-body wave function over the Pauli string configurations. The sampling is achieved with a novel MPS technique, which enables to compute each sample in an efficient way with a computational cost $O(N\chi^3)$.
We benchmark our method over randomly generated magic states, as well as in the ground-state of the quantum Ising chain. Exploiting the extremely favourable scaling, we easily have access to the non-equilibrium dynamics of the SREs after a quantum quench.
\end{abstract}

\maketitle

\paragraph{Introduction. --}
The concept of quantum advantage~\cite{Preskill_2012,Preskill_2018} is based on the idea that the intrinsic exponential complexity of a quantum system can be exploited to overcome the limitations of a classical computation, possibly providing ways to solve NP problems efficiently~\cite{Feynman_1982, kitaev2002classical,365700}. Physicists agree to identify the {\it entanglement} as a fundamental feature accounting for this complexity, thus making necessary to exploit it proficiently in any quantum computation. Indeed, quantifying the entanglement of a many-body system is a long standing argument of research~\cite{Vedral_1997}
and many different measurements (e.g.\ purity, entanglement entropy, negativity, mutual information, entanglement capacity, etc.) are nowadays well known and studied \cite{RevModPhys.80.517,PasqualeCalabrese_2004,de_Boer_2019,Dalmonte_2022}. Nevertheless, entanglement is not the sole {\it resource} which need to be quantified in order to discriminate between 
easy and hard to simulate quantum states.
Indeed, it turns out that there exist several states encoding an extensive amount of entanglement which can still be simulated efficiently on a classical computer. Those states belong to the class of {\it stabilizer states}~\cite{https://doi.org/10.48550/arxiv.quant-ph/9705052}, which by definition are the quantum states that can be prepared by means of only Clifford unitaries from the computational basis state $\ket{0\dots0}$~\cite{Gottesman_1998,Aaronson_2004,https://doi.org/10.48550/arxiv.1711.07848, https://doi.org/10.48550/arxiv.2007.05992, https://doi.org/10.48550/arxiv.quant-ph/9807006, Kitaev_2002}.
Clifford group represent a class of unitary transformations that acts as the normalizer of the $n$-qubits Pauli group, i.e.\ mapping Pauli strings into Pauli strings~\cite{https://doi.org/10.48550/arxiv.quant-ph/9807006, nielsen_chuang_2010}. Because of this underlying structure, any stabilizer state can be represented classically in a compact form, and any Clifford manipulation can be performed efficiently operating in this representation~\cite{Gottesman_1998,Aaronson_2004}. 
As a consequence, in order to quantify the hardness of simulating a quantum state, regardless its entanglement content, it is crucial to define a quantity which accounts for the 
amount of non-Clifford operations needed to prepare a state~\cite{Howard_2014,Seddon_2021}.
This quantity has been dubbed {\it nonstabilizerness} or {\it quantum magic}. It turns out that nonstabilizerness is also related to the emergence of quantum chaos~\cite{Leone_2021,GotoChaosbyMagic}. Several measures of magic have been proposed so far in quantum information theory 
\cite{PRXQuantum.3.020333, Howard_2017}, 
as for instance the Robustness of Magic~\cite{Howard_2017}. 
Nevertheless they are typically hard to compute~\cite{Heinrich_2019}.
As a matter of fact, quantifying nonstabilizerness beyond a few qubits remains a major challenge.
Recently, an efficient procedure based on Bell measurements over two copies of a state has been proposed~\cite{Haug_2023}.
Meanwhile, the {\it Stabilizer Rényi Entropies} (SREs) were introduced in Ref.~\cite{Leone_2022} as a possible way of quantifiying the magic content of a quantum state.
Interestingly, SREs allow the evaluation of the magic stored in the ground state of the paradigmatic transverse field Ising chain~\cite{Oliviero_2022}.
However, since they depend explicitly on the expectation value of all possible Pauli strings, the cost of computing SREs for a generic state scales exponentially with the number of qubits.
Nevertheless when the $N$-qubits state admits a Matrix Product State (MPS) representation with finite bond dimension $\chi$, 
the SREs can be computed as the norm of a ``$2n$-replica'' MPS with effective bond dimension $\chi^{2n}$, where $n$ (integer) represents the Rényi index~\cite{Piroli_2023}. Unfortunately,
such norm can be computed at a cost $O(N\chi^{6n})$, thus having an unfavorable scaling with the bond dimension. Indeed, for any practical purpose, this makes that approach unfeasible for $n>2$~\footnote{Only for $n=2$, it is possible to exploit additional symmetries, further reducing the computational cost to $O(\chi^4)$.}. 

To overcome such limitations we propose a breakthrough method which exploits the probabilistic nature of the SREs. The Algorithm relies on a novel and efficient MPS sampling in the Pauli basis, which acts in a way reminiscent of some well-established MPS techniques~\cite{Stoudenmire_2010, Ferris_2012}. By sampling over $\nsamp$ Pauli strings realizations, we are able to estimate the SREs with a computational cost which scales as $O(\nsamp N \chi^{3})$. 
We first benchmark our approach over a set of random realization of MPS states with large bond dimension.   
We then study the magic in the ground-state of the quantum Ising chain, showing a prefect agreement with the free-fermions calculation.
Finally, we use our method to compute for the first time the non-equilibrium dynamics of the SREs after a quench. We consider the Ising model with or without a longitudinal field
and show how the confinement of the excitations~\cite{Kormos_2016}, which hugely affect the entanglement dynamics, may play a role also in the time-evolution of the SREs. \\

\paragraph{Preliminaries. ---}
Let us consider a quantum system consisting of $N$ qubits.
We identify the Pauli matrices by $\{\sigma^{\alpha}\}_{\alpha=0}^{3}$, 
with $\sigma^{0} = \mathbb{1}$, and with $\pmb{\sigma} = \prod_{j=1}^{N}\sigma_{j} \in \mathcal{P}_{N}$ a generic $N-$qubits Pauli strings where $\mathcal{P}_{N} = \{ \sigma^0, \sigma^1, \sigma^2, \sigma^3 \}^{\otimes N}$. 
For a pure normalised state $\rho = |\psi\rangle \langle \psi|$, the SREs  \cite{Leone_2022} are given by 
\begin{equation}\label{eq:sre}
    M_{n} ( \rho ) = \frac{1}{1 - n} 
    \log 
     \sum_{\pmb{\sigma}\in\mathcal{P}_{N}} \frac{1}{2^N} \Trace[\rho \, \pmb{\sigma}]^{2 n}.
\end{equation}
To understand the relation with usual Rényi entropies, let us consider the non-negative real-valued function 
$\Pi_{\rho}(\pmb{\sigma}) = \frac{1}{2^N} \Trace[\rho \,\pmb{\sigma}]^{2}$. We have indeed
\begin{equation}
\sum_{\pmb{\sigma}\in\mathcal{P}_{N}} \Pi_{\rho}(\pmb{\sigma}) =
\Trace \left[ \rho \, \sum_{\pmb{\sigma}\in\mathcal{P}_{N}} \frac{1}{2^N}
\Trace[\rho \, \pmb{\sigma}]  \pmb{\sigma} \right] 
= \Trace[\rho^2] = 1 ,
\end{equation}
where we used the unique decomposition  
$\rho = 2^{-N}\sum_{\pmb{\sigma} }  \Trace[\rho \, \pmb{\sigma}] \, \pmb{\sigma}$, in terms of the Pauli matrices,
which are a complete orthonormal basis with respect to the 
scalar product $\Trace[\pmb{\sigma}\,\pmb{\sigma'}] = 2^{N}\delta_{\pmb{\sigma}\pmb{\sigma'}}$. We can thus interpret $\Pi_{\rho}(\pmb{\sigma})$ as a probability distribution on the set of Pauli strings. Therefore
$ M_{n} ( \rho ) = (1-n)^{-1}
\log  \sum_{\pmb{\sigma}\in\mathcal{P}_{N}} \Pi_{\rho}(\pmb{\sigma})^{n}  - N \log 2
$,
apart from a constant, does coincides with the $n$-Rényi entropy of the distribution $\Pi_{\rho}(\pmb{\sigma})$, and it reduces to the Shannon entropy
$
M_{1}(\rho) = -\sum_{\pmb\sigma\in\mathcal{P}_{N}}
\Pi_{\rho}(\pmb{\sigma}) \log \Pi_{\rho}(\pmb{\sigma}) - N \log(2)
$
for $n\to 1$. Let us mention that, the definition of $M_{n}(\rho)$ can be easily extended to arbitrary (non-pure) states $\rho$ by normalizing the probability with the purity $\Trace[\rho^2]\neq 1$, thus redefining  
$ \Pi_{\rho}(\pmb{\sigma}) = \frac{1}{2^{N}} \Trace[\rho \, \pmb{\sigma}]^{2} / \Trace[\rho^2] $. It has been shown that SREs have the following properties \cite{Leone_2022}, accordingly being a good measure of magic: i) $M_n$ vanishes for stabilizer states whereas is positive for other states; ii) are invariant under Clifford unitaries; iii) are additive.
Moreover, they grow extensively with the system size $N$, thus making possible to define a density of magic as $m_n = M_n / N$~\cite{Piroli_2023}.
Recently, a violation of monotonicity for the SREs with $0 \leq n < 2$ has been reported for systems undergoing measurements in the computational basis~\cite{haug2023stabilizer}.

Computing the SREs in Eq.~(\ref{eq:sre})
requires the evaluation of 
the expectation value of
a generic power $\Pi_{\rho}(\pmb{\sigma})^{n-1}$  (or   $\log\Pi_{\rho}(\pmb{\sigma})$ for $n=1$) over the probability distribution $\Pi_{\rho}(\pmb{\sigma})$ itself. This suggests a natural way to estimate the SREs, based on a sampling from $\Pi_{\rho}(\pmb{\sigma})$.\\


\paragraph{Conditional sampling. --}

The task of sampling from the set of the Pauli strings $\pmb{\sigma}$, which has size $D=4^N$, may appear as exponentially hard. To overcome this difficulty we rewrite the full probability in terms of conditional and prior (or marginal) probabilities as
\be\label{eq:chain_prob}
\Pi_{\rho}(\pmb{\sigma})
= \pi_{\rho}(\sigma_1)
\pi_{\rho}(\sigma_2|\sigma_1)
\cdots
\pi_{\rho}(\sigma_N|\sigma_1\cdots\sigma_{N-1})
\ee
where 
$
\pi_{\rho}(\sigma_j|\sigma_{1}\cdots \sigma_{j-1}) =
\frac{\pi_{\rho}(\sigma_1\cdots \sigma_j)}
{\pi_{\rho}(\sigma_1\cdots \sigma_{j-1})}
$
is the probability that the Pauli matrix $\sigma_{j}$ occurs at position $j$ given that the string $\sigma_1\cdots\sigma_{j-1}$ has already occurred at positions $1 \dots j-1$, no matter the occurrences in the rest of the system (i.e.\ marginalising over all possible Pauli strings for the reaming qubits $j+1 \dots N$). Specifically, one has
$
\pi_{\rho}(\sigma_1\cdots \sigma_j) 
= \sum_{\pmb\sigma \in \mathcal{P}_{N-j}}
 \frac{1}{2^N} \Trace[\rho \, \sigma_{1}\cdots\sigma_{j}\pmb{\sigma}]^{2}
$.
In other terms, the conditional probability at the step $j$, i.e.
$\pi_{\rho}(\sigma_j|\sigma_{1}\cdots \sigma_{j-1})$,
can be thought as the probability $\pi_{\rho_{j-1}}(\sigma_{j})$ of getting $\sigma_{j}$ in the partially projected state
\be\label{eq:rho_j}
\rho_{j-1}  \equiv  
\frac{\rho|_{\sigma_1\cdots\sigma_{j-1}}}{\pi_{\rho}(\sigma_1\cdots \sigma_{j-1})^{1/2}}
\ee
where we have defined the state 
$
\rho|_{\sigma_1\cdots\sigma_{j-1}}
\equiv
2^{-N}\sum_{\pmb{\sigma}\in\mathcal{P}_{N-j+1}}  
\Trace[\rho \, \sigma_{1}\cdots \sigma_{j-1} \pmb{\sigma}]
\,\sigma_{1}\cdots\sigma_{j-1}\pmb{\sigma}
$
where, in the Pauli matrices decomposition of $\rho$, we are only keeping the contribution with fixed $\sigma_1\cdots\sigma_{j-1}$. Notice that such state is not normalised, however $\Trace[\rho_{j-1}^2]=1$, and the probability that the remaining string $\pmb{\sigma}\in\mathcal{P}_{N-j+1}$ occurs is exactly given by $\pi_{\rho}(\pmb{\sigma}|\sigma_{1}\cdots\sigma_{j-1})$.
From the definition in Eq.~(\ref{eq:rho_j}), we can easily get the recursive relation 
$
\rho_{j} = 
\pi_{\rho_{j-1}}(\sigma_{j})^{-1/2}
\rho_{j-1}|_{\sigma_{j}}
$.
Thanks to that, we can generate the outcomes (and the probabilities of that outcomes) by iterating over each single qubits, and sampling each local Pauli matrix according to the conditional probabilities. Once a local outcome occurs, the state is updated accordingly, and the iteration proceeds until all qubits are sampled. At the end of this procedure, as a direct result of the chain rule in Eq.~(\ref{eq:chain_prob}), we generated configurations $\pmb\sigma$ with probability $\Pi_{\rho}(\pmb{\sigma})$.    
Of course, in order for this method to be computationally affordable, we need an efficient way of: {\it (i)} evaluating the conditional probabilities; {\it (ii)} updating the state according to the local outcome. In the following Section we show that these conditions are met whenever the state admits an MPS representation.\\

\begin{figure}[t!]
\includegraphics[width=0.8\linewidth]{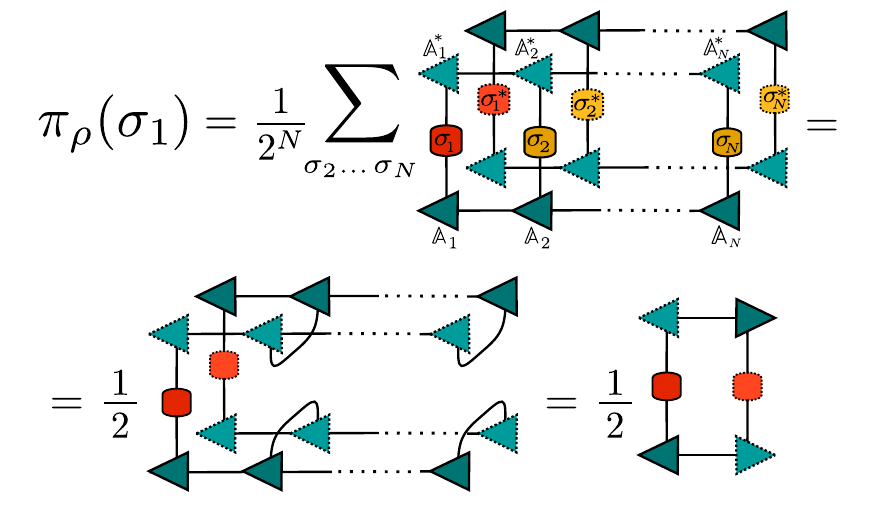}
\caption{MPS evaluation of the marginal probability $\pi_{\rho}(\sigma_1)$. 
Dotted lighter shapes represent conjugate tensors. Contractions over the auxiliary indices can be easily carried out thanks to the property in Eq.~(\ref{eq:Lambda}), together with the right-normalization of the $\mathbb{A}_i$ tensors. 
\label{fig:2} }
\end{figure}

\begin{algorithm}[H]
\caption{Pauli sampling from MPS}\label{alg:QA}
\begin{flushleft}
\hspace*{\algorithmicindent} \textbf{Input}: an MPS $\ket{\psi}$ of size $N$ 
\end{flushleft}
\begin{algorithmic}[1]
\State Put the MPS in right-normalized form. 
\State Initialize $\mathbb{L}=(1)$ and $\Pi = 1$ (see Fig.\ref{fig:1} $a)$)
\For{($i=1$, $i=N$, $i++$)}
     \State Compute the probabilities
     $\pi(\alpha)=\pi_{\rho}(\sigma^{\alpha}|\sigma_1\cdots\sigma_{i-1})$
     \indent for $\alpha \in \{ 0,1,2,3 \}$ as in Fig.\ref{fig:1} $b)$.
     \State  Generate a random value of $\alpha$ according to $\pi(\alpha)$
     \State Set $\sigma_i = \sigma^{\alpha}$, update $\Pi \rightarrow \Pi \cdot \pi(\alpha)$ 
     \State Update $\mathbb{L}$  as in Fig.\ref{fig:1} $c)$.
\EndFor
\end{algorithmic}
\begin{flushleft}
\hspace*{\algorithmicindent} \textbf{Output}: a Pauli string $\pmb{\sigma}$ and the probability $\Pi(\pmb{\sigma})$
\end{flushleft}
\end{algorithm}

\paragraph{MPS iterative algorithm. --}
 Let us consider a pure state $|\psi\rangle$ represented in  
 the following MPS form~\cite{schollwoeck2011,Silvi_2019,PhysRevLett.91.147902}
 \begin{equation}
 |\psi\rangle = \sum_{s_1,s_2,\dots,s_N} \mathbb{A}^{s_1}_{1}\mathbb{A}^{s_2}_{2}\cdots\mathbb{A}^{s_N}_{N} |s_1,s_2,\dots,s_N \rangle,
 \end{equation}
 with $\mathbb{A}^{s_{j}}_{j}$ being $\chi\times\chi$ matrices,
 except at the left (right) boundary where $\mathbb{A}^{s_{1}}_{1}$
 ($\mathbb{A}^{s_{N}}_{N}$) is a $1\times\chi$ ($\chi\times1$) row (column) vector.  Here $|s_{j}\rangle \in \{|0\rangle, |1\rangle\}$ is a local computational basis. The state is assumed right-normalised, namely
 $\sum_{s_{j}} \mathbb{A}^{s_{j}}_{j} (\mathbb{A}^{s_{j}}_{j})^{\dag} = \Id$. 
 Following the conditional sampling prescription described in the previous section, we start from the first term of the expansion in Eq.~(\ref{eq:chain_prob}). This can be written as
\begin{align}\label{eq:pi1}
    \begin{split}
    \pi_{\rho}(\sigma_1) 
    &= \frac{1}{2^N} \sum_{\pmb{\sigma} \in \mathcal{P}_{N-1}} \braket{\psi| \sigma_1  \pmb{\sigma} |\psi} \braket{\psi^*| \sigma_1^* \pmb{\sigma}^* |\psi^*} \, ,
    \end{split}
\end{align}
where we used the fact that the Pauli matrices are hermitian. 
In terms of the operators $\Lambda_{\sigma_i} = \frac{1}{2} \sigma_i \otimes \sigma_i^*$ and $\Lambda_i = \frac{1}{2} \sum_{\sigma_i} \big( \sigma_i \otimes \sigma_i^* \big)$, each acting on the local Hilbert space given by a spin and its replica, the previous
equation reads
$\pi_{\rho}(\sigma_1) = 
\big[ \bra{\psi} \otimes \bra{\psi^*} \big] \Lambda_{\sigma_1}  \Lambda_{2}\cdots 
\Lambda_{N}  \big[ \ket{\psi} \otimes \ket{\psi^*} \big]$.
Now, the following property can easily be proven  
\begin{equation}\label{eq:Lambda}
\big[ \bra{s_i'}\otimes\bra{r_i'} \big] \Lambda_{i} \big[\ket{s_i}\otimes\ket{r_i}\big] = \delta_{s_i',r_i'} \delta_{s_i,r_i} \, ,
\end{equation}
meaning that $\Lambda_{i}$ is just two copies of the identity operator connecting the spin $|s_i\rangle$ and its replica (whose local computational basis is now indicated as $|r_{i}\rangle \in \{|0\rangle, |1\rangle\}$). 
Using Eq.~(\ref{eq:Lambda}) together with the right-normalization of the MPS, the computation of Eq.~(\ref{eq:pi1}) reduces in the following local tensor contraction 
\begin{equation}
\pi_{\rho}(\sigma_1) =
\frac{1}{2} 
\sum_{{s_1,s_1',r_1,r_1'}}
(\mathbb{A}^{s_1'}_{1})^* \mathbb{A}^{r_1'}_{1}
(\sigma_1)_{s_1' s_1} (\sigma_1^*)_{r_1' r_1} \mathbb{A}^{s_1}_{1} (\mathbb{A}^{r_1}_{1})^* \, \, ,
\end{equation}
which is represented in Fig.~\ref{fig:2} by means of the standard Tensor Network graphical notation \cite{schollwoeck2011, Silvi_2019}. 

\begin{figure}[t!]
\includegraphics[width=0.8\linewidth]{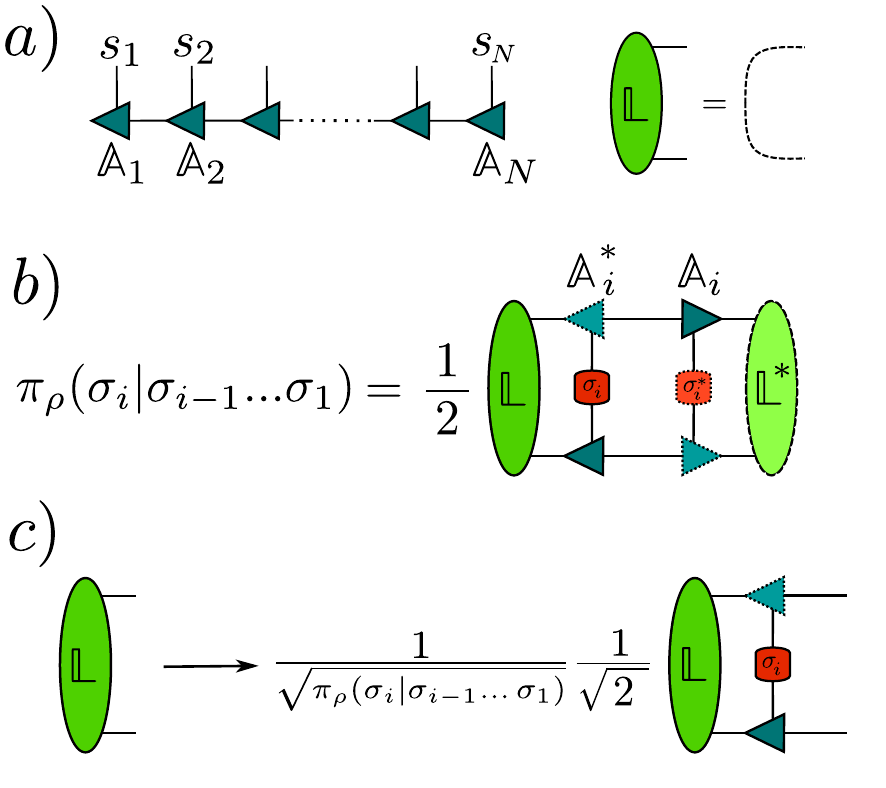}
\caption{The iterative sampling Algorithm \ref{alg:QA}. \label{fig:1} }
\end{figure}

After evaluating $\pi_{\rho}(\sigma_1)$ for each $\{ \sigma^0, \sigma^1, \sigma^2, \sigma^3 \}$ one can extract a sample from this distribution, thus obtaining the first element of the Pauli string. 
The information about the partially projected state
Eq.~(\ref{eq:rho_j}) is encoded in an effective environment matrix $\mathbb{L} = \frac{1}{\sqrt{2 \, \pi_{\rho}(\sigma_1)}} \sum_{{s_1,s_1'}} (\mathbb{A}^{s_1'}_{1})^* (\sigma_1)_{s_1' s_1} \mathbb{A}^{s_1}_{1}$. 
The calculation of the next terms of Eq.~(\ref{eq:chain_prob}) 
and the extraction of the remaining $\sigma_i$ proceeds following the same line. The full sampling recipe is summarized in the Algorithm~\ref{alg:QA},
and graphically supported in Fig.~\ref{fig:1}. 
Our approach can be generalized to estimate 
$M_n(\rho)$ in the case in which $\rho$ is the reduced density matrix 
describing the rightmost $N$ qubits embedded in a larger pure MPS state.
It is easy to show that this operation would only affect the initialization of the matrix $\mathbb{L}$ (which is set to $(1)$ for a pure state), where in the general case 
 $\mathbb{L} = \mathbb{\Lambda}^2/\sqrt{\Trace(\mathbb{\Lambda}^4)}$
in terms of the Schmidt eigenvalues $\mathbb{\Lambda}$.\\

\paragraph{Sampling error. --} In this section we discuss the statistical errors associated with the proposed sampling algorithm, and their scaling with the system size $N$. Let us first consider the case of estimating the $n-$SRE, with $n>1$. As we saw, the estimation of $\Pn = \sum_{\pmb{\sigma}\in\mathcal{P}_{N}} \Pi_{\rho}(\pmb{\sigma})^{n}$ is achieved by a statistical average over the samples $\{ \pmb{\sigma}_{\mu} \}_{\mu = 1}^{\nsamp}$, that means using the estimator
\begin{equation}\label{eq:estimator}
\PnEx = \frac{1}{\nsamp} \sum_{\mu=1}^{\nsamp} \Pi_{\rho}(\pmb{\sigma}_{\mu})^{n-1} \, . 
\end{equation}
Afterwards, we evaluate the density of magic as $\tilde{m}_n = \big(N(1-n)\big)^{-1} \log \PnEx - \log 2$. Notice that $\PnEx$ is an unbiased estimator of $q_n$, since $\overline{\PnEx} = \Pn$ ($\overline{ \phantom{a} }$ indicating the average over the uncorrelated samples, each distributed according to $\Pi_{\rho}(\pmb{\sigma})$). 
The fluctuations of $\PnEx$ are characterized by its variance, which can be easily evaluated as $\Var [ \PnEx ] = \Var [ \Pi_{\rho}^{n-1} ] / \nsamp$. For every $n>1$, one has $\Var [ \Pi_{\rho}^{n-1} ] < 1$ and thus we can upper bound the variance of the estimator obtaining $\Var [ \PnEx ] < \text{const.} / \nsamp$,
where const.\ is a constant of $o(1)$, whose value is independent of the size $D=4^N$ of the support of $\Pi_{\rho}(\pmb{\sigma})$. This means that the statistical error on $\PnEx$ can be reduced arbitrarily by increasing the number of samples, no matter the system size $N$. However,  
since the uncertainty on $\tilde{m}_n$ propagates (at first order) as $\delta \tilde{m}_n \propto \delta \PnEx / \PnEx$ and both $\PnEx$, $\delta \PnEx$ are exponentially vanishing with $N$ for typical probability distributions, 
$(\delta \tilde{m}_n)^2 \sim 
\frac{1}{\nsamp} \Var[ \Pi_{\rho}^{n-1} ] / (\overline{\Pi_{\rho}^{n-1}})^2$ is generally 
exponentially \textit{increasing} with $N$
\footnote{This because $\overline{\Pi_{\rho}^{2(n-1)}} / (\overline{\Pi_{\rho}^{n-1}})^2 \geq 1$}. Nevertheless, 
for physical relevant states the estimation error $\delta \tilde{m}_n$ is always under control for reasonable values of $\nsamp$ (see next Section and Supplementary Materials for further details).
For $n=1$ we evaluate $\PnOne = \sum_{\pmb{\sigma}\in\mathcal{P}_{N}} \Pi_{\rho}(\pmb{\sigma}) \log \Pi_{\rho}(\pmb{\sigma})$ via the estimator
\begin{equation}
\PnExOne = \frac{1}{\nsamp} \sum_{\mu=1}^{\nsamp} \log \Pi_{\rho}(\pmb{\sigma}_{\mu}) \, . 
\end{equation}
We have $\Var[\PnExOne] = \Var[\log \Pi_{\rho}] / \nsamp$ and thus we are interested in giving an upper bound for $\Var[\log \Pi_{\rho}]$. Several works, e.g. Ref.~\cite{de_Boer_2019}, establish that $\Var[\log \Pi_{\rho}] \leq \frac{1}{4} \log^2(D) + 1$. Thus, in our case, $\Var[\PnExOne] \lesssim N^2 \log^2(2) / \nsamp$ meaning that \textit{in the worst scenario} the number of samples has to scale as $N^2$ to reach a given accuracy in the estimation. Finally, let us observe that, after having generated the samples and the corresponding probabilities, one can in principle devise better ways of post-processing the data (for instance via improved estimators).\\

\begin{figure}[t!]
\centering
\includegraphics[width=0.85\linewidth]{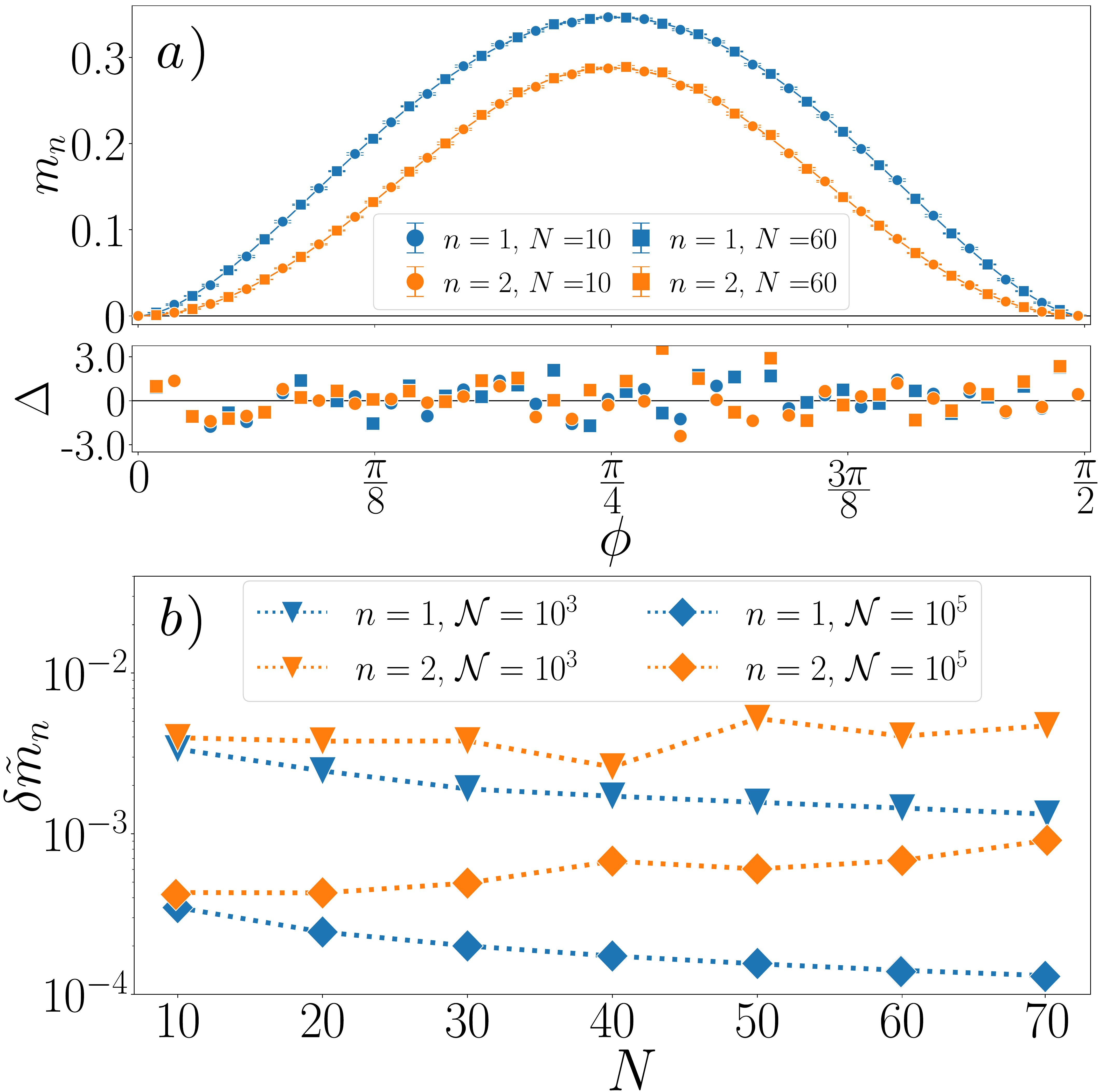}
\caption{$a)$ Density of magic of $\ket{\psi} = U_{\mathcal{C}} \ket{T_{\phi}}^{\otimes N}$ for $N=10,60$, $\nsamp = 10^4$ and Rényi index $n=1,2$. In the lower strip we show the deviation from the analytical value $\Delta = (m_n - \tilde{m}_n) / \delta \tilde{m}_n$, $\tilde{m}_n$ being our estimation and $\delta \tilde{m}_n$ the propagated statistical error. $b)$ The error $\delta \tilde{m}_n$ as a function of the system size $N$ for fixed $\nsamp = 10^3, 10^5$ and $\phi \simeq \pi/4$. \label{fig:randomclifford} }
\end{figure}

\begin{figure}[t!]
\centering
\includegraphics[width=1.03\linewidth]{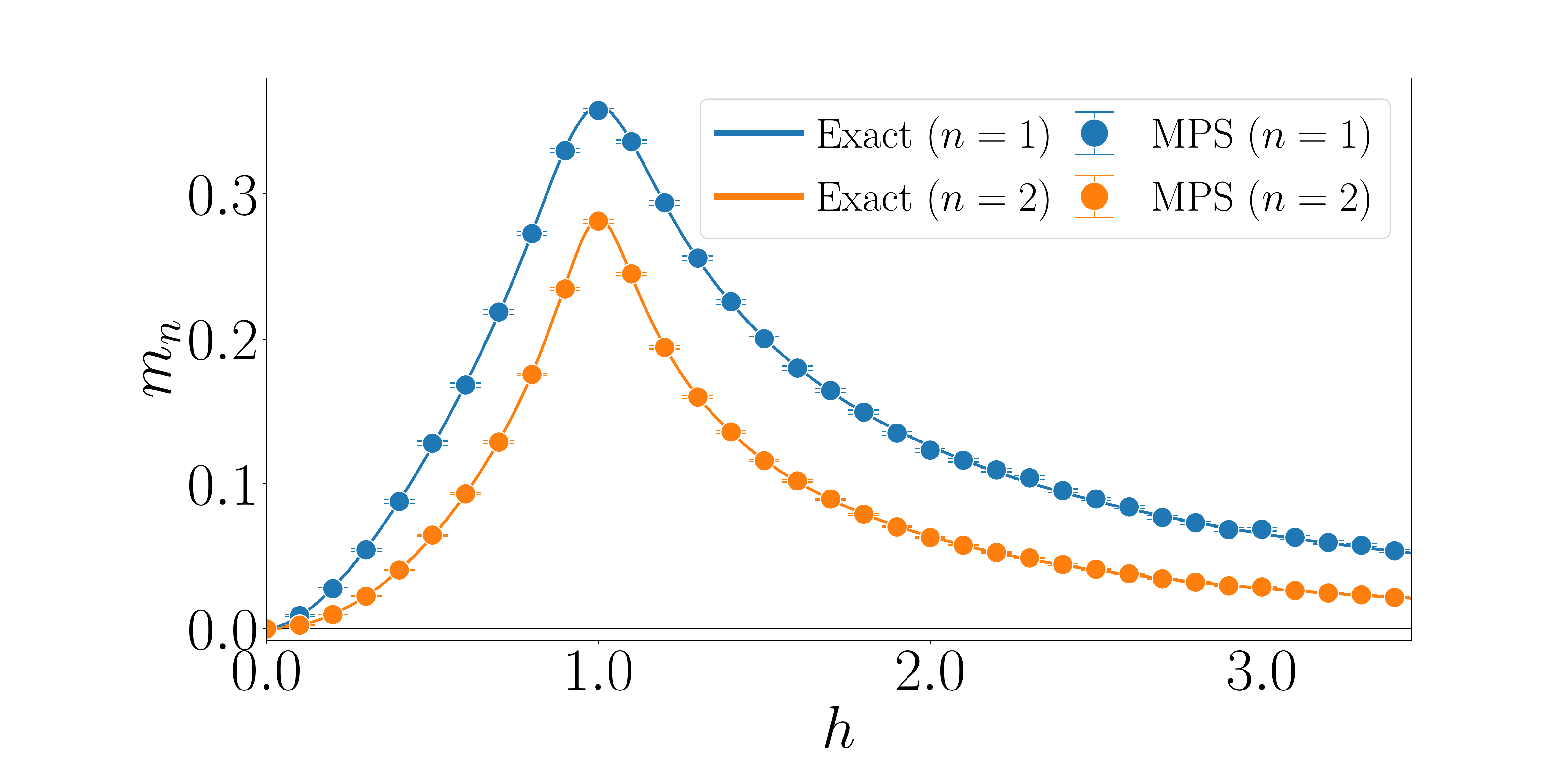}
\caption{Density of magic of the Ising ground state ($g=0$) with periodic boundary conditions, for a system of size $N=14$ and Rényi index $n=1,2$. Exact results obtained in the free fermions representation~\cite{Oliviero_2022} are compared with MPS sampling ($\nsamp = 10^4$). \label{fig:ising} }
\end{figure}

\paragraph{Numerical experiments. --} As a first benchmark of our algorithm, we considered the $T$-state $\ket{T_{\phi}} = ( \ket{0} + e^{i \phi} \ket{1} ) / \sqrt{2}$, with $\phi$ ranging in $[0,\pi/2]$. A straightforward calculation yields to $M_2(\ket{T_{\phi}}\bra{T_{\phi}}) = - \log [(1 + \cos^4\phi + \sin^4\phi)/2]$, and $M_1(\ket{T_{\phi}}\bra{T_{\phi}}) = -  \cos^2 \phi \, \log ( | \cos \phi | ) - \sin^2 \phi \, \log (|\sin \phi |)$. Both the quantities vanish for $\phi=0,\pi/2$, while they have a maximum for $\phi=\pi/4$. In our experiment, we firstly initialize the system in the product state $\ket{\psi_0} = \ket{T_{\phi}}^{\otimes N}$, which is an MPS of bond dimension $\chi = 1$. Afterwards, we apply a random unitary Clifford circuit $U_{\mathcal{C}}$ of depth $N$. In each layer, we randomly choose a sequence of one-qubits or two-qubits
gates extracted from the Clifford generators $\{\Id,S,H,\text{CNOT}\}$. The final MPS $\ket{\psi} = U_{\mathcal{C}} \ket{\psi_0}$ has a larger bond dimension $\chi \gg 1$, whereas its content of magic is the same of $\ket{\psi_0}$, since the  magic is invariant under Clifford group. Thanks to the additivity of the SREs, the density of magic $m_n(\ket{\psi}) = M_n(\ket{\psi}) / N$ is equivalent to the magic of a single $T-$state. We apply our sampling algorithm on the MPS $\ket{\psi}$, obtaining the estimation $\tilde{m}_n$. Results are shown in Fig.~\ref{fig:randomclifford}, for $n=1,2$ and systems of size between $N=10$ and $N=70$. Notice that for $N=70$, the bond dimension of $\ket{\psi}$ grows up to $\chi=128$, depending on the particular arrangement of the Clifford layers. Values of $\chi$ of this order would be extremely challenging to target with previously known methods~\cite{Piroli_2023}, whereas our approach takes only $\approx O(0.1) \, $sec/sample on a single node simulation. Notice that the sampling can be easily parallelized, provided that the MPS is stored in multiple independent copies. All the data points are in agreement with theoretical predictions within three error bars (see Fig.~\ref{fig:randomclifford} $a)$). Moreover, a scaling of the statistical error $\delta \tilde{m}_n$ with $N$ at fixed value of $\nsamp$ suggests that the fluctuations do not grow significantly with the system size, 
even though in principle we might have expected them to increase exponentially with $N$ for $n=2$.

Afterwards, we consider the quantum Ising model $H = - \sum_{i} \sigma^x_i \sigma^x_{i+1} - h \sum_{i} \sigma^z_i 
- g \sum_{i} \sigma^x_i $. For $g=0$, this Hamiltonian can be easily mapped into a model of free fermions \cite{https://doi.org/10.48550/arxiv.2009.09208, Calabrese_2012}, thus allowing the evaluation of
the SREs in terms of $\sim 4^N$ determinants of matrices involving fermionic correlators \cite{Oliviero_2022}. In Fig.~\ref{fig:ising}, we compare exact results for $m_n$ ($n=1,2$) obtained in the fermionic representation with MPS estimations, for a system of size $N=14$. In the MPS approach we find the ground state using standard one-site DMRG ($\chi=32$). MPS data are in perfect agreement with the exact values, within small error bars. 

\begin{figure}[t!]
\centering
\includegraphics[width=1.03\linewidth]{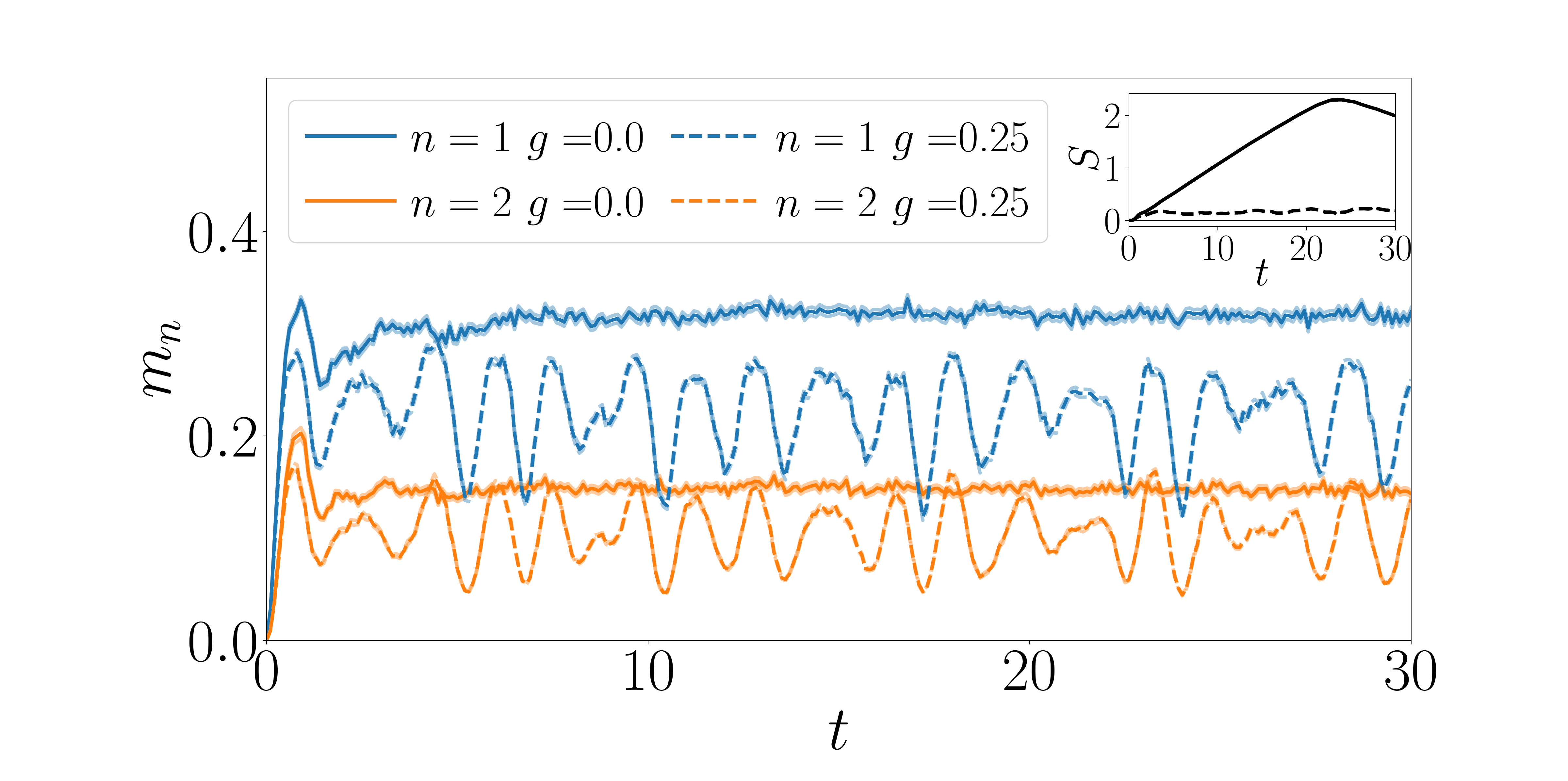}
\caption{Density of magic after a quantum quench in the transverse and longitudinal field Ising model ($N=40$). The system is prepared in the ferromagnetic state $\ket{+ ... +}$ and quenched with parameters $h=0.5, g=0.0$ (solid line), $h=0.5, g=0.25$ (dotted line). Magic estimation is obtained with $\nsamp = 10^3$ samples and pale lines represent the corresponding statistical uncertainty. Subplot: half-chain entanglement entropy. \label{fig:isinglong} }
\end{figure}

Finally, we use our algorithm to estimate the dynamics of the magic density during an out-of-equilibrium protocol. In particular, we prepare the system in the fully polarized state $\ket{\psi(0)} = \ket{+ \dots +}$, where $\ket{+} = (\ket{0} + \ket{1})/\sqrt{2}$ is the eigenstate of $\sigma^x$ with eigenvalue $+1$, and we consider the time-evolution generated by the Ising hamiltonian i.e.\ $\ket{\psi(t)} = e^{-i H t} \ket{\psi(0)}$. We set the transverse and longitudinal fields respectively to $h=0.5$ and $g=0, \, 0.25$. The latter value corresponds to a phase in which the system is known to exhibit a dynamical confinement of the excitations~\cite{Kormos_2016, https://doi.org/10.48550/arxiv.2212.00533}, whereas in the free case ($g=0$) the quasiparticles give rise to a light cone spreading of correlations \cite{PhysRevLett.96.136801}.  
We use the TEBD approach to compute the time evolution of the post-quench MPS~\cite{PhysRevLett.93.040502,schollwoeck2011}, with bond-dimension up to $\chi=128$. Results are shown in Fig.~\ref{fig:isinglong} for $N=40$. For $g=0$, the magic density seems to saturate rapidly to a stationary value, although the half-chain entanglement entropy $S = -\Trace [\varrho_{N/2}\log\varrho_{N/2}]$, is still growing linearly with the time $t$ as expected (see the subplot). In the confined phase $g=0.25$, the magic $m_{n}$ exhibits much larger oscillations around a slightly lower stationary value, whereas the entanglement is strongly suppressed and it approaches to a very low saturation value.\\

\paragraph{Conclusions. --}
We have shown that a relatively new measure of quantum nonstabilizerness, the Stabilizer Rényi Entropies~\cite{Leone_2022}, can be estimated efficiently in the MPS framework, via a perfect sampling in the space of Pauli strings operators. Our estimation neither suffers from the exponential growth of the size of the many-body Hilbert space, nor shows an unfavorable scaling with the MPS bond dimension.
As a matter of fact, we are able to consider either equilibrium or non-equilibrium wave-functions with MPS bond-dimension up to values that were out of reach by any of the previously proposed methods for evaluating the nonstabilizerness. Specifically, we applied our method to evaluate the amount of magic generated after a quench in the quantum Ising chain, and its sensitivity to the presence of confinement of excitations. Although we mainly focused on pure MPS, our algorithm can be easily adapted to non-pure states obtained from an MPS tracing out a subsystem consisting of the first or last qubits.

Our approach pave the way to novel extensive numerical studies of the quantum magic, possibly providing new interesting characterizations of the quantum phases of matter, in and out-of-equilibrium. In addition, our new Pauli sampling technique for the MPS can be used to address crucial problems in quantum many-body theory, as for instance the operator scrambling.\\

We thank Lorenzo Piroli for significant remarks on our manuscript and Alessandro Laio for useful suggestions about sampling strategies.
We are particularly grateful to Titas Chanda, Marcello Dalmonte, Alioscia Hamma, and Emanuele Tirrito for collaborations on topics connected with this work, and for inspiring discussions. 
This work was supported by the PNRR MUR project PE0000023-NQSTI (M.C.). \\

\bibliography{bib}

\end{document}


\onecolumngrid
\appendix
\appendixtitleon
\appendixtitletocon

\begin{appendices}
\section{SUPPLEMENTARY MATERIALS}

\section{Considerations on the sampling error and sampling from factorized states}

As discussed in the main text, for Rényi index $n>1$, our approach relies on the following estimation 
\begin{equation}
\PnEx = \frac{1}{\nsamp} \sum_{\mu=1}^{\nsamp} \Pi_{\rho}(\pmb{\sigma}_{\mu})^{n-1} \qquad \tilde{m}_n = \frac{1}{(1-n) N} \log \PnEx - \log 2 \, ,
\end{equation}
of the quantities 
\begin{equation}
\Pn = \sum_{\pmb{\sigma}\in\mathcal{P}_{N}} \Pi_{\rho}(\pmb{\sigma})^{n}  \qquad m_n = \frac{1}{(1-n) N} \log \Pn - \log 2 \, .
\end{equation}
The error of the estimation $\tilde{m}_n$ is given by
\begin{align}\label{eq:tayloreq}
\begin{split}
\tilde{m}_n - m_n  =  \frac{1}{(1-n) N} \log( 1 \pm \sqrt{\zn}) \qquad \quad \zn = \frac{\big( \PnEx - \Pn \big)^2 }{\Pn^2} \, ,
\end{split}
\end{align}
where $\pm$ corresponds to ${\rm sgn}(\PnEx/\Pn-1)$. The goodness of the estimation depends on the value taken by the random variable $\zn$. To study the statistical properties of $\zn$, let us preliminarily define the $k-$th central moment of a generic random variable $X$ as 
\begin{equation}
    \mu_{X, k} = \overline{(X - \overline{X})^k} \, , 
\end{equation}
such that $\mu_{X, 1} = 0$, $\mu_{X, 2} = \Var[X]$. Since $\PnEx$ is defined as the sample mean of $\Pi_{\rho}(\pmb{\sigma})^{n-1}$, one can easily obtain $\overline{\PnEx} = \Pn$ and also 
\begin{align}
\begin{split}
\mu_{\PnEx, 2} &= \frac{1}{\nsamp} \mu_{\Pi^{n-1}, 2} \\ 
\mu_{\PnEx, 4} &= \frac{3}{\nsamp^2} (\mu_{\Pi^{n-1}, 2})^2 + \frac{1}{\nsamp^3} \bigg( \mu_{\Pi^{n-1}, 4} - (\mu_{\Pi^{n-1}, 2})^2 \bigg) \, . 
 \end{split}
\end{align}
Furthermore 
\begin{align}
\begin{split}
 \overline{\zn} & = \frac{\mu_{\PnEx, 2}}{q_n^2}  \qquad \qquad 
 \Var[\zn] =  \overline{\zn}^2 \bigg( \frac{\mu_{\PnEx, 4}}{\mu_{\PnEx, 2}^2}   - 1 \bigg) \, .
 \end{split}
\end{align}
Now, we have $\overline{\zn} = \overline{(\PnEx)^2} / \Pn^2 - 1$, where both $\overline{(\PnEx)^2}, \Pn^2$ are expected to be exponentially small in the system size $N$ for a fixed value of $\nsamp$. Instead, the ratio $\overline{(\PnEx)^2}/\Pn^2 >1$, and hence also $\overline{\zn}$, is in general exponentially large.  However, this does not necessary mean that also the typical value taken by $\zn$ is also exponentially large. Indeed, let us notice that the fluctuations of $\zn$ around $\overline{\zn}$ are also exponentially increasing and even larger than $\overline{\zn}$, since $\Var[\zn]/\overline{\zn}^2 \simeq   \mu_{\PnEx, 4} / \mu_{\PnEx, 2}^2$ and the ratio $\mu_{\PnEx, 4} / \mu_{\PnEx, 2}^2 > 1$ is exponentially large.\\

\begin{figure}
\subfigure[]{\includegraphics[width=8.7cm]{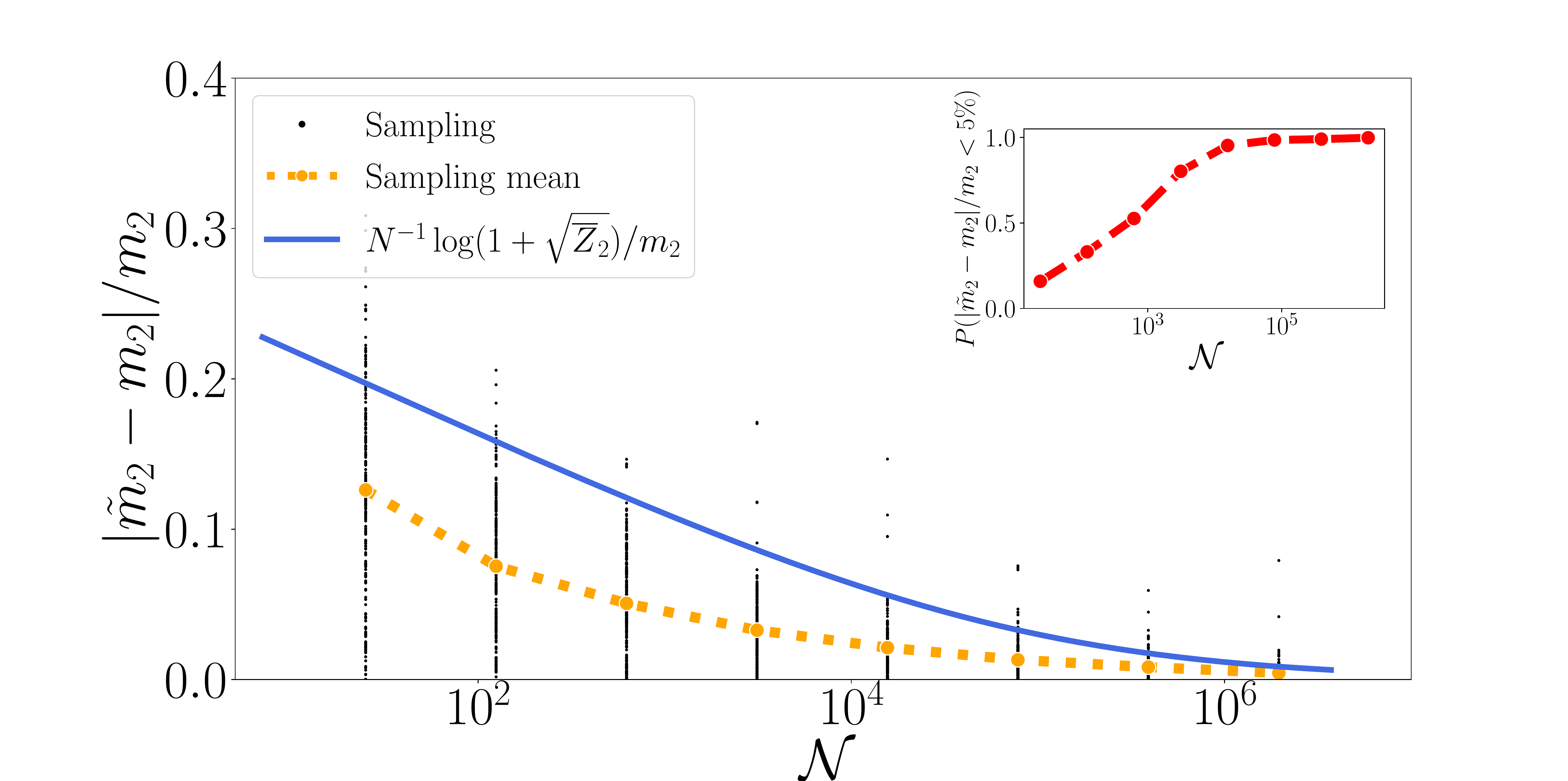}}
\subfigure[]{\includegraphics[width=8.7cm]{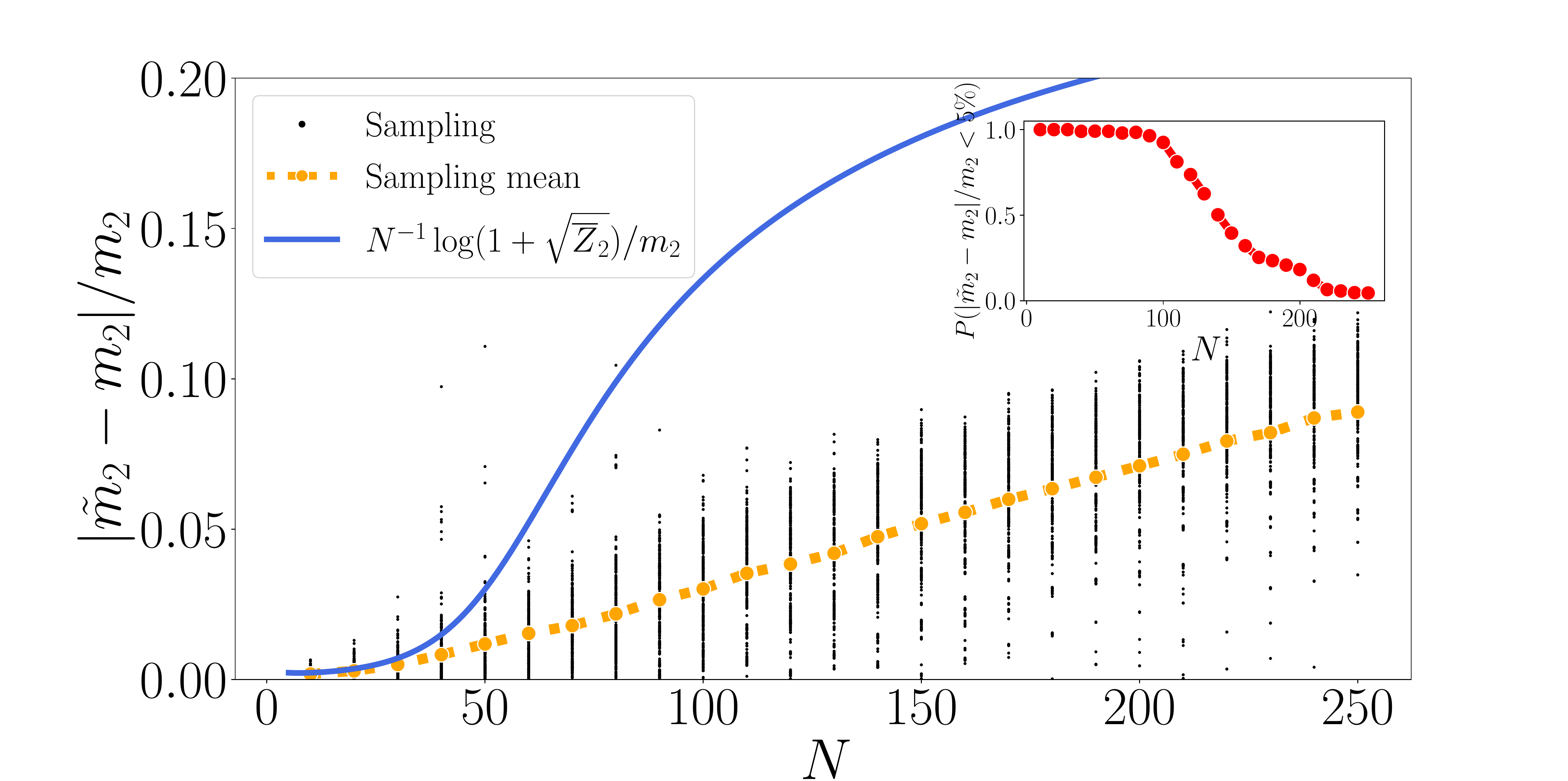}}
\caption{Relative error $|\tilde{m}_2 - m_2|/m_2$ of the SRE estimation for the product state $(\rho_0)^{\otimes N}$, with $\rho_0$ as in Eq.~\ref{eq:prodsate} ($p_1=p_2=1/3$). $(a)$ $|\tilde{m}_2 - m_2|/m_2$ as a function of $\nsamp$, for fixed $N=50$. Black dots represent different estimations, each obtained with the same number of samples $\nsamp$. Subplot: empirical value of the probability $P(|\tilde{m}_2 - m_2|/m_2 < 5\%)$. $(b)$ $|\tilde{m}_2 - m_2|/m_2$ as a function of $N$ for fixed $\nsamp=10^5$. Black dots represent different estimations,each obtained with the same number of samples $\nsamp=10^5$. Subplot: empirical value of the probability $P(|\tilde{m}_2 - m_2|/m_2 < 5\%)$.
\label{fig:prodstatenew} }
\end{figure}

As a simple case study, we now consider the class of product states $\ket{\psi} = \ket{\psi_0}^{\otimes N}$. We can write the local density matrix in the Pauli basis as 
\begin{equation}\label{eq:prodsate}
  \rho_0 = \ket{\psi_0}\bra{\psi_0} = \frac{1}{2} \Id + \sum_{\alpha=1}^3 \bigg(\frac{p_{\alpha}}{2}\bigg)^{1/2} \frac{\sigma^{\alpha}}{\sqrt{2}} \, ,
\end{equation}
where $p_{\alpha}/2$ are the probabilities to sample $\sigma^{\alpha}$. To ensure $\Trace[\rho_0^2]=1$, one needs $p_1 + p_2 + p_3 =1$. Now, we can write the probability $\Pi_{\rho}(\pmb{\sigma})$ in the following way
\begin{equation}\label{eq:probmult}
     \Pi_{\rho}(\pmb{\sigma}) = \frac{1}{2^{N}} \Trace[(\rho_0)^{\otimes N} \, \pmb{\sigma}]^{2} = \frac{1}{2^{N}} \prod_{i=1}^N \Trace[\rho_0 \, \sigma_i]^2 = \bigg(\frac{1}{2} \bigg)^{N_{0}} \bigg(\frac{p_1}{2} \bigg)^{N_{1}}  \bigg(\frac{p_2}{2} \bigg)^{N_{2}} \bigg(\frac{1 - p_1 - p_2}{2} \bigg)^{N_{3}} \, ,
\end{equation}
where $N_{\alpha}$ is the number of times $\sigma^{\alpha}$ appears inside $\pmb{\sigma}$ (thus, $N_0 + N_1 + N_2 +N_3 =N$). The sampling is therefore described by the following multinomial probability distribution 
\begin{equation}
     \Pi_{\rho}(N_0, N_1, N_2, N_3) =  \binom{N}{N_0, N_1, N_2, N_3} \bigg(\frac{1}{2} \bigg)^{N_{0}} \bigg(\frac{p_1}{2} \bigg)^{N_{1}}  \bigg(\frac{p_2}{2} \bigg)^{N_{2}} \bigg(\frac{1 - p_1 - p_2}{2} \bigg)^{N_{3}} \, .
\end{equation}
A simple explicit calculation for $n=2$ yields to 
\begin{equation}
     \overline{\znd} = \frac{1}{\nsamp} \bigg[ \bigg(2 \frac{1 + p_1^3  + p_2^3 + (1 - p_1 - p_2)^3}{\big(1 + p_1^2  + p_2^2 + (1 - p_1 - p_2)^2 \big)^2 }\bigg)^N - 1 \bigg] \, ,
\end{equation}
which, as expected, increase exponentially with $N$ at fixed $\nsamp$. It is straightforward to verify that the expression inside  parenthesis has a maximum for $p_1 = p_2 = 1/3$ and for these values one has $\overline{\znd} = \frac{1}{\nsamp} \big( (5/4)^N - 1 \big)$. 
This is expected to be the worst scenario for the sampling Algorithm.
For the same values of $p_1, p_2$, one can also get
$\Var[\znd] \simeq \frac{1}{\nsamp^3} (41/16)^N$ at the leading order for $N \gg 1$. 
Furthermore, $q_2 = 1/3^{N}$ and $m_2 = \log(3/2)$. In Figure~\ref{fig:prodstatenew}, we plot data obtained by performing a perfect sampling of the product state $(\rho_0)^{\otimes N}$ (with $p_1 = p_2 = 1/3$) to estimate $m_2$. In Figure~\ref{fig:prodstatenew} $(a)$, we analyze the relative estimation error $|\tilde{m}_2 - m_2|/m_2$ as a function of $\nsamp$, for fixed $N=50$. The sampling has been iterated $10^3$ times for each fixed values of $\nsamp$ in order to collect statistics on the values of $|\tilde{m}_2 - m_2|/m_2$ (see black points). Blue line represents the analytical value obtained by replacing $\znd$ with $\overline{\znd}$ in Eq.~\ref{eq:tayloreq}.Since $\log$ is a concave function, this value is an upper bound on the mean error, i.e.\ $\log( 1 + \sqrt{\overline{\znd}}) \geq \overline{\log( 1 + \sqrt{\znd})}$. In our numerical experiment, we observe that in most of the cases the error is greatly lower than this value. For example, for $\nsamp = 10^3$, we have $N^{-1} \log( 1 + \sqrt{\overline{\znd}}) / m_2 \simeq 0.11$, whereas we empirically find $\overline{|\tilde{m}_2 - m_2|} / m_2 \approx 0.04$ (orange line) and $|\tilde{m}_2 - m_2| < 5 \% \, m_2$ in the $\approx 60 \%$ of the cases (see subplot). In Figure~\ref{fig:prodstatenew} $(b)$, the number of samples is fixed to $\nsamp = 10^5$, and we show the relative error $|\tilde{m}_2 - m_2|/m_2$ for different system sizes $N$. Once again, we iterate the sampling $10^3$ times (black points). As before, the error is typically significantly lower than the expected (blue line). For the largest system size considered, $N=250$, the average relative error (orange line) is lower than the $10\%$.

\end{appendices}